\documentclass [aps,prl,twocolumn,showpacs] {revtex4}
\usepackage[final]{graphics}
\usepackage{amssymb}
\usepackage{amsfonts}
\usepackage{epsfig}

\begin{document}
\date{\today}
\title{Free energy variational approach for 
the classical anisotropic $XY$ model in a crystal field}
\author{L. M. Castro$^1$, A. S. T. Pires$^2$ and J. A. Plascak$^2$ }
\affiliation{$^1$Departamento de Ci\^encias Exatas, Universidade Estadual do 
Sudoeste da Bahia, Estrada do Bem Querer Km04, CP 95, 45083-900 Vitoria da 
Conquista, BA - Brazil \\
 $^2$Departamento de F\'\i sica, Instituto de Ci\^encias Exatas,
 Universidade Federal de Minas Gerais, C.P. 702, 
 30123-970 Belo Horizonte, MG - Brazil}

\begin{abstract}

A variational approach for the free energy is used to study the
three-dimensional anisotropic 
$XY$ model in the presence of a crystal field. The magnetization and the phase
diagrams as a function of the parameters of the Hamiltonian are obtained. 
Some limiting results for isotropic $XY$ and       
planar rotator models in two and three dimensions are analyzed and compared to 
previous results obtained from analytical approximations as well as from 
those obtained from more reliable approaches such as series expansion and 
Monte Carlo simulations. It is also shown that from this general variational 
approach  some simple assumptions can drastically simplify the 
self-consistent implicit equations. 
The validity of the low temperature region of this approach is 
analyzed and compared to Monte Carlo results as well. 

\end{abstract} 

\maketitle
\section{1. Introduction}

Ising and Heisenberg models are spin Hamiltonians which have been
originally proposed in the study of magnetism in pure and diluted materials
\cite{kobe,stinch,belan,wolf}. They are widely studied and employed in both 
classical
and quantum  contexts. However, a great interest has also been devoted
to the $XY$ model due to the richness of its phase diagram and the character
of its transition \cite{be,kt}. In particular, the XY model in two dimensions
is of prime interest in the field of statistical mechanics
because of its intriguing and unusual phase transition.
Recently, the study of the $XY$ model has 
been motivated not only due to its applicability in describing
real magnetic materials (for instance, it has been quite recently shown that
the phase diagram of the compound RbFe(MoO$_4$)$_2$ is in good agreement
with the theoretical predictions for a two-dimensional XY classical
model \cite{svistov}) but as well as in condensed  matter systems, 
for example liquid crystals \cite{liu} and 
superconductors \cite{pierson,choi,franz}, among others. 

The one-dimensional version of the quantum spin-1/2 $XY$ model has been
exactly solved by Lieb, Shultz and Mattis \cite{lms}. 
Although the quantum nature of real magnets cannot be forgotten, 
the study of classical models continues to be an important subject of 
research. The two-dimensional classical version of the $XY$ model
has been treated by Berezinskii and Kosterlitz and Thouless \cite{be,kt} and
they showed that the vortices and anti-vortices play a central role in the 
thermodynamic behavior of the model. For higher dimensions 
this model (as well as a planar rotator version)
has been studied through approximate analytical techniques
\cite{beth,tonico3,tonico1,tonico2} and Monte Carlo simulations \cite{landau}.
In addition, classical models in general, and in particular the XY model, 
have become
very popular recently in the context of quantum phase transitions, where a
d-dimensional quantum model at $T=0$ is transformed into a classical $d+1$
dimensional one \cite{sach,lonz}.

The effort for a better theoretical understanding of the high
temperature superconductors has also lead to an increase in 
treating anisotropic $XY$ type models \cite{franz,min,shenoy,eric},
specially its two-dimensional version. 
However, despite the layered structure of these systems,
the high temperature superconductors are not strictly two
dimensional. For this reason, inter layer interaction should be important
in describing their thermodynamic properties \cite{franz}, 
as well as the inclusion of possible crystal field interactions.

In this work we study the classical anisotropic $XY$ model with a crystal 
field interaction described by the following Hamiltonian
\begin{eqnarray}
\label{ham} {\cal H}&=&-J\sum_{ \langle \vec{r},
\vec{r}^{'}\rangle}\{S_{\vec{r}}^{x} S_{\vec{r}^{'}}^{x}+
S_{\vec{r}}^{y}S_{\vec{r}^{'}}^{y}\} 
\nonumber\\
&-&J_{z}\sum_{ \langle \vec{r},
\vec{r}^{'}\rangle}\{S_{\vec{r}}^{x} S_{\vec{r}^{'}}^{x}+
S_{\vec{r}}^{y}S_{\vec{r}^{'}}^{y}\}+
D\sum_{\vec{r}}(S_{\vec{r}}^{z})^{2}~,
\end{eqnarray}
where $J$ is the exchange interaction between spins in the layers parallel to
the $xy$ plane and $J_z$ the exchange interaction between spins in
different adjacent layers. $D$ is the crystal field and $S_{\vec{r}}^{\alpha}$
are the $\alpha=x,y,z$ components of a classical spin $|\vec{S}_{\vec{r}}|=1$.
The first sum runs on nearest neighbor spins $\langle \vec{r},
\vec{r}^{'}\rangle$ within the layers and the second sum  on
nearest neighbor spins $\langle \vec{r}, \vec{r}^{'}\rangle$ between layers
(in order to avoid confusion here and in the following expressions,
 the subscript of the exchange interaction or the variational parameter 
in front of the sums will define whether the spins belong to the same layer
or to adjacent layers). The last sum is made over the entire $N$ spins on 
the simple cubic lattice. 

The Hamiltonian (\ref{ham}) can be written in a more convenient form by
means of a polar representation for the spins \cite{beth,tonico3}
\begin{eqnarray}
\label{polar}
\vec{S_{\vec{r}}} &=& (S_{\vec{r}}^{x}, S_{\vec{r}}^{y}, S_{\vec{r}}^{z}) = 
({\sin \theta_{\vec{r}}} {\cos \phi_{\vec{r}}},
{\sin \theta_{\vec{r}}} {\sin \phi_{\vec{r}}}, S_{\vec{r}}^{z}), 
\nonumber\\
\vec{S_{\vec{r}}}&=&({\sqrt{1- {(S_{\vec{r}}^{z}})^2}\cos \phi_{\vec{r}}},
 \sqrt{1- {(S_{\vec{r}}^{z}})^2}{\sin \phi_{\vec{r}}}, 
S_{\vec{r}}^{z})~,
\end{eqnarray}
where $\theta_{\vec r}$ and $\phi_{\vec r}$ are the spherical angles of the 
spin at the site $\vec r$. In this representation
the Hamiltonian given by Eq. (\ref{ham}) takes the form
\begin{eqnarray}
\label{hplanar} {\cal H}&=&-\frac{J}{2}\sum_{  \vec{r}, \vec{a}}
\sqrt{1- {(S_{\vec{r}}^{z}})^{2}} \sqrt{1-
{(S_{\vec{r}+\vec{a}}^{z}})^{2}}\cos
(\phi_{\vec{r}+\vec{a}}-\phi_{\vec{r}})
  \nonumber\\
&-&\frac{J_{z}}{2}\sum_{ \vec{r}, \vec{c}} \sqrt{1-
{(S_{\vec{r}}^{z}})^{2}} \sqrt{1-
{(S_{\vec{r}+\vec{c}}^{z}})^{2}}\cos
(\phi_{\vec{r}+\vec{c}}-\phi_{\vec{r}}) 
\nonumber\\
&+&D\sum_{\vec{r}}(S_{\vec{r}}^{z})^{2},
\end{eqnarray}
where  $\vec a$ labels the four nearest-neighbor sites of $\vec r$ 
in the $xy$ plane and $\vec c$ the two nearest-neighbor sites of 
$\vec r$ along the $z$ direction.

The above model has been treated according to the self-consistent harmonic
approximation (SCHA) in the case $D=0$ for the three-dimensional model
\cite{tonico2} as well as the planar rotator model \cite{tonico1}
(in this case there is no $z$ spin components). The approach
in \cite{tonico2}, however, only
considers the cases $J_z\approx J$ and $J_z<<J$, that is, quasi-isotropic 
case or weak layer coupling and no general treatment has been done for any
value of $0\le J_z\le J$.

In what concerns the theoretical approach, it is well known that
the SCHA has been extensively used in the literature 
\cite{franz,tonico1,tonico2,eric,ari1,ari2,spis,wood,lozo,fish1,
fish2,ari3,darcy}. 
It consists of replacing the original Hamiltonian (\ref{hplanar}) by
a harmonic one by expanding the corresponding cosines to the second order
with effective exchange constants that take into account the nonlinearities
of the interactions. The effective couplings are then chosen to minimize the
corresponding free energy of the system. 
In the present work we use a variational method based on Bogoliubov inequality 
for the free energy to study the model Hamiltonian (\ref{ham})
in order to obtain its low temperature thermodynamic properties.
This procedure is the same as the SCHA employed in previous works. 
Some simplifications are also suggested  in order to 
easier handle the awkward self-consistent implicit equations.

The plan of the paper is as follows. In the next section, we present the
analytical procedure according to the variational approach for the free
energy in order to obtain the thermodynamic properties of the model. 
In section 3, the
numerical results are presented, where the role played by the anisotropy in 
the presence of a crystal field is shown to be for itself  relevant (for 
instance, in the context of high temperature superconductors). Additional 
assumptions are also suggested which can make the
implicit self-consistent equations easier to be handled. Some concluding 
remarks are discussed in section 4.

\section{2. Variational Approach for the free energy}

The present variational approach is based on the Bogoliubov inequality for
the free energy
\begin{equation} \label{bi}
F\leq F_{0}+\langle {\cal H}-{\cal H}_0(\gamma)
\rangle_{0}\equiv \Phi(\gamma) ,
\end{equation}
where ${\cal H}$ is the Hamiltonian in study given by Eq. (\ref{ham})
or (\ref{hplanar}), ${\cal H}_0(\gamma)$ is a trial Hamiltonian which can be 
exactly solved and depends on variational parameters $\gamma$. $F$ is the free 
energy of the system described by $\cal H$, $F_o$ is the free energy of the
trial Hamiltonian ${\cal H}_o$, and the thermal average $<... >_0$ is taken
over the ensemble defined by ${\cal H}_0$. The
approximate free energy is given by the minimum of $\Phi(\gamma)$ with
relation to $\gamma$, that is, $F\equiv \Phi_{min}(\gamma)$.

In general, the trial Hamiltonian should resemble, in some aspects, the
one under study. In this case, ${\cal H}_0$ can be chosen  as a sum of two
parts
\begin{eqnarray}
\label{Ho} {\cal H}_{o}= {\cal H}_{o}^\phi + {\cal H}_{o}^z,
\end{eqnarray}
in such a way that the first part is a kind of a planar Hamiltonian 
\begin{eqnarray}
\label{Hopb} {\cal H}_{o}^\phi=\frac{\gamma}{4}\sum_{
\vec{r}, \vec{a}}(\phi_{\vec{r}+\vec{a}}-\phi_{\vec{r}})^{2}
+\frac{\gamma_{z}}{4}\sum_{ \vec{r},
\vec{c}}(\phi_{\vec{r}+\vec{c}}-\phi_{\vec{r}})^{2}
\end{eqnarray}
and the second term is an axial Hamiltonian 
\begin{eqnarray}
\label{Hoab}
{\cal H}_{o}^z=
(D+2J+J_{z})\sum_{
\vec{r}}(S_{\vec{r}}^{z})^{2} ,
\end{eqnarray}
where $\gamma$ and $\gamma_z$ stand for the variational parameters.
This harmonic choice for ${\cal H}_0$
is also motivated by the fact that at low temperatures the
angle differences $|\phi_{\vec{r}+\vec{a}}-\phi_{\vec{r}}|<<1$ and
$|\phi_{\vec{r}+\vec{c}}-\phi_{\vec{r}}|<<1$ so the cosines in Eq.
(\ref{hplanar}) can be expanded up to second order to give the terms
$(\phi_{\vec{r}+\vec{a}}-\phi_{\vec{r}})^2$ and
$(\phi_{\vec{r}+\vec{c}}-\phi_{\vec{r}})^2$ in ${\cal H}_0^\phi$.
Since in this case there is
a negligible variation in the $z$ components of the spins, we can further
assume that $S_{\vec r}^z\approx S_{\vec r + \vec a}^z$ and
$S_{\vec r}^z\approx S_{\vec r + \vec c}^z$,  so the square roots in Eq.
(\ref{hplanar}) are eliminated and we end up to a term $D+2J+J_z$ in
the corresponding axial term. Thus, it turns out that this approach will
be valid only in the low temperature region.

Obtaining the averages regarding the trial Hamiltonian requires 
diagonalizing (\ref{Ho}). 
The planar Hamiltonian ${\cal H}_0^\phi$ can be diagonalized in the reciprocal
space through the Fourier transform of $\phi_{\vec{r}}$
\begin{equation}
\label{ft}
\begin{array}{ccc}
\phi_{\vec{r}}=\frac{1}{\sqrt{N}}\sum_{\vec{q}}e^{-i\vec{q}\cdot\vec{r}}
\phi_{\vec{q}} ,
\end{array}
\end{equation}
with the inverse transform $\phi_{\vec{q}}$ given by
\begin{equation}
\label{ift} \phi_{\vec{q}}=\frac{1}{\sqrt{N}}\sum_{
\vec{r}}e^{i\vec{q}\cdot\vec{r}} \phi_{\vec{r}}.
\end{equation}
Using the transformations (\ref{ft}) in Eq. (\ref{Hopb}) and the fact that
the system is translationally invariant we get
\begin{eqnarray}
\label{Hopfp} {\cal H}_{o}^{\phi}&=&
\frac{1}{2}\sum_{\vec{q}}\{\gamma\sum_{\vec{a}}
(1-e^{-i\vec{q}\vec{a}})
\phi_{\vec{q}}\phi_{\vec{-q}}\nonumber \\
&+&\gamma_z\sum_{\vec{c}}
(1-e^{-i\vec{q}\vec{c}})\phi_{\vec{q}}\phi_{\vec{-q}}\}.
\end{eqnarray}
Summing now over the vectors ${\vec a}$ in the $xy$ plane and $\vec c$
along the $z$ direction and rearranging terms
we obtain the diagonal form of the planar trial Hamiltonian
\begin{equation}
\label{Hopf} {\cal
H}_{o}^{\phi}=\sum_{\vec{q}}(\gamma_{q}+\gamma_{qz})\vert \phi_{q}
\vert^{2},
\end{equation}
where $\gamma_{q}=\gamma(2-\cos q_{x}a -\cos q_{y}a)$,
$\gamma_{qz}=\gamma_{z}(1-\cos q_{z}c)$, $a=|\vec a|$,
 $c=|\vec c|$ and $\vert \phi_{q}
\vert^{2}= \phi_{\vec{q}}\phi_{-\vec{q}}$.

The diagonalizing of the axial term of the harmonic Hamiltonian (\ref{Hoab})
is obtained by introducing the corresponding Fourier transform of the $z$
component of the spins $S_{\vec{r}}^{z}$
\begin{eqnarray} \label{fts}
S_{\vec{r}}^{z}=\frac{1}{\sqrt{N}}\displaystyle\sum_{\vec{q}}e^{-i\vec{q}
\cdot\vec{r}}S_{\vec{q}}^{z} ,
\end{eqnarray}
and the inverse transform $S_{\vec{q}}^{z}$ given by
\begin{eqnarray}
\label{ifts}
S_{\vec{q}}^{z}=\frac{1}{\sqrt{N}}\sum_{  \vec{q}}e^{i\vec{q}\cdot\vec{r}}
S_{\vec{r}}^{z}.
\end{eqnarray}
Applying (\ref{fts}) to Eq. (\ref{Hoab}) one gets
\begin{eqnarray}
\label{Hoaq}
{{\cal H}_{o}^{z}}=
(D+2J+J_{z})\sum_{  \vec{q}}
|S_{\vec{q}}^{z}|^2,
\end{eqnarray}
where  $\vert S_{\vec{q}}^{z}\vert^{2}=S_{\vec{q}}^{z}S_{-\vec{q}}^{z}$.

The partition function ${\cal Z}_0$ can be computed by first noting that
the planar and the axial parts of the harmonic
Hamiltonian are independent so
\begin{equation}
{\cal Z}_0=Tre^{-\beta{\cal H}_0}=Tre^{
-\beta({\cal H}_{o}^\phi + {\cal H}_{o}^z)}={\cal Z}_0^\phi{\cal Z}_0^z.
\end{equation}
Since both ${\cal H}_0^\phi$ and ${\cal H}_0^z$ are quadratic in their
variables one has
\begin{equation}
\label{Zop} {\cal Z}_{o}^{\phi}=Tre^{-\beta {\cal H}_{o}^\phi}=
\prod_{\vec{q}}\left[\frac{\pi}{\beta (
\gamma_{q}+\gamma_{qz})}\right]^{\frac{1}{2}}
\end{equation}
and
\begin{equation}
\label{Zoa} {\cal Z}_{o}^{z}=Tre^{-\beta {\cal H}_{o}^z}=
\prod_{\vec{q}}\left[\frac{\pi}{\beta \Omega}\right]^{\frac{1}{2}},
\end{equation}
where $\Omega=D+2J+J_{z}$.
The free energy $F_0$ is then given by
\begin{eqnarray}
\label{F0} F_0=-
\frac{k_{B}T}{2}\sum_{\vec{q}}\ln{\frac{\pi}{\beta (
\gamma_{q}+\gamma_{qz})}}-\frac{k_{B}T}{2}N\ln{\frac{\pi}{\beta
\Omega}}
\end{eqnarray}

The mean value $<{\cal H}_0>_0$ can be evaluated by using the equipartition
theorem resulting in
\begin{equation}
\label{NKT} \langle {\cal
H}_{0}\rangle_{0}=\frac{Nk_{B}T}{2}+\frac{Nk_{B}T}{2}=Nk_{B}T.
\end{equation}

On the other hand, the mean value of $<{\cal H}>_0$ is not so straightforward
computed. It can be written as
\begin{eqnarray}
\label{Hm1}
\langle{\cal H}\rangle_{0}= \hspace{7cm} \nonumber \\
- \frac{J}{2}\sum_{ \vec{r},
\vec{a}}\langle \sqrt{1- (S_{\vec{r}}^{z})^{2}} \sqrt{1-
(S_{\vec{r}+\vec{a}}^{z}})^{2}\rangle_{0}\langle
\cos (\phi_{\vec{r}+\vec{a}}-\phi_{\vec{r}})\rangle_{0}\nonumber\\
- \frac{J_{z}}{2}\sum_{  \vec{r}, \vec{c}}\langle
\sqrt{1- (S_{\vec{r}}^{z}})^{2} \sqrt{1-
(S_{\vec{r}+\vec{c}}^{z}})^{2}\rangle_{0}\langle
\cos (\phi_{\vec{r}+\vec{c}}-\phi_{\vec{r}})\rangle_{0}
\nonumber \\
+D\sum_{\vec{r}}\langle(S_{\vec{r}}^{z})^{2}\rangle_{0},\hspace{5.5cm}
\end{eqnarray}
where in the  first two sums the mean value of the product of the planar and
axial terms regarding the trial Hamiltonian has been factorized and
\begin{equation}
\label{sm2}
\langle(S_{\vec{r}}^{z})^{2}\rangle_{0}=\frac{k_{B}T}{
2(D+2J+J_z)}
\end{equation}
is the out-of-plane spin fluctuation. 

For the
Gaussian variables $(\phi_{\vec{r}+\vec{a}}-\phi_{\vec{r}})$ we can write
\begin{equation}
\label{gauss}
 \langle \cos
(\phi_{\vec{r}+\vec{a}}-\phi_{\vec{r}}) \rangle_{0}
=e^{-\frac{1}{2}\langle (\phi_{\vec{r}+\vec{a}}-\phi_{\vec{r}})^2
\rangle_{0}},
\end{equation}
where the mean value appearing in the exponential is given by
\begin{eqnarray}
\label{difangaf}  \langle (\phi_{\vec{r}+\vec{a}}-\phi_{\vec{r}})^2 \rangle_0
=\frac{2}{N}
\sum_{  \vec{q}}(1-\lambda_{\vec{q}})
\langle |\phi_{\vec{q}}|^2\rangle_0,
\end{eqnarray}
where $\lambda_{{q}}=\frac{1}{2}(\cos q_{x}a +\cos q_{y}a)$
and
\begin{equation}
\label{phiq2}
\langle\vert \phi_{q}\vert^{2}\rangle_{0}=\frac{k_{B}T}{
2(\gamma_{q}+\gamma_{qz})}.
\end{equation}
Similarly, for the Gaussian variable $(\phi_{\vec{r}+\vec{c}}-\phi_{\vec{r}})$
we have
\[
 \langle \cos
(\phi_{\vec{r}+\vec{c}}-\phi_{\vec{r}}) \rangle_{0}
=e^{-\frac{1}{2}\langle (\phi_{\vec{r}+\vec{c}}-\phi_{\vec{r}})^2
\rangle_{0}},
\]
and
\begin{equation}
\label{difangcf}
\langle(\phi_{\vec{r}+\vec{c}}-\phi_{\vec{r}})^2\rangle_0 =\frac{2}{N}
\sum_{  \vec{q}}(1-\lambda_{\vec{q}}^{z})
\langle |\phi_{\vec{q}}|^2 \rangle_0,
\end{equation}
where $\lambda_{{qz}}=\cos q_{z}c$. In this way, Eq. (\ref{Hm1})
assumes the form
\begin{eqnarray}
\label{origmed2} \langle {\cal H}\rangle_{0}= -
\frac{J}{2}\sum_{ \vec{r}, \vec{a}} (1- \langle
(S_{\vec{r}}^{z})^{2}\rangle_{0}) e^{-\frac{1}{N}
\sum_{\vec{q}}(1-\lambda_{q})\langle\vert \phi_{q}
\vert^{2}\rangle_{0}} \nonumber\\
- \frac{J_{z}}{2}\sum_{  \vec{r}, \vec{c}}(1- \langle
(S_{\vec{r}}^{z})^{2}\rangle_{0})e^{-\frac{1}{N}\sum_{\vec{q}}(1-\lambda_{qz})
\langle\vert\phi_{q}\vert^{2}\rangle_{0}}\nonumber\\
+\sum_{\vec{q}}D\langle \vert
S_{\vec{q}}^{z}\vert^{2}\rangle_{0},\hspace{4cm}
\end{eqnarray}
where we have used an additional assumption that
$S_{\vec r}^z\approx S_{\vec r + \vec a}^z$ and
$S_{\vec r}^z\approx S_{\vec r + \vec c}^z$. 
The last term can again
be computed from the equipartition theorem and as the terms in the
sums do not depend on the respective indexes we have
\begin{eqnarray}
\label{H0f}
{\langle {\cal H}\rangle_{0}}= \hspace{7cm} \nonumber \\
- 2JN(1- \langle
(S_{\vec{r}}^{z})^{2}\rangle_{0})
e^{-\frac{1}{2N}\sum_{\vec{q}}(2-(\cos{q_{x}a} +
\cos{q_{y}a}))\langle\vert \phi_{q}
\vert^{2}\rangle_{0}} \nonumber\\
-J_{z}N(1- \langle
(S_{\vec{r}}^{z})^{2}\rangle_{0})e^{-\frac{1}{N}\sum_{\vec{q}}
(1-\cos{q_{z}c})\langle\vert
\phi_{q}\vert^{2}\rangle_{0}}+\frac{NDk_BT}{2\Omega}.
\end{eqnarray}
The right hand side of Eq. (\ref{bi}) is then written as
\begin{eqnarray}
\Phi(\gamma ,\gamma_{z})= -
\frac{k_{B}T}{2}\sum_{\vec{q}}\ln{\frac{\pi}{\beta (
\gamma_{q}+\gamma_{qz})}}-\frac{k_{B}T}{2}\sum_{\vec{q}}\ln{\frac{\pi}{\beta
\Omega}}\nonumber\\
- 2JN(1- \langle
(S_{\vec{r}}^{z})^{2}\rangle_{0})
e^{-\frac{1}{2N}\sum_{\vec{q}}(2-(\cos{q_{x}a} +
\cos{q_{y}a}))\langle\vert \phi_{q}
\vert^{2}\rangle_{0}}\nonumber\\
-\mbox{} J_{z}N(1- \langle
(S_{\vec{r}}^{z})^{2}\rangle_{0})e^{-\frac{1}{N}\sum_{\vec{q}}(1-
\cos{q_{z}c})\langle\vert
\phi_{q}\vert^{2}\rangle_{0}}
\nonumber\\
-(2-\frac{D}{\Omega})\frac{NK_BT}{2}.
\label{phi}
\end{eqnarray}

Minimizing the above equation with respect to the variational parameters gives
an upper bound limit for the free energy. The variational parameters are
determined from the conditions
\begin{equation}
\label{dg}
\frac{\partial \Phi(\gamma ,\gamma_{z})}{\partial
\gamma}=0~ (a)\mbox{~~~~~and~~~~~} \frac{\partial \Phi(\gamma ,\gamma_{z})}
{\partial
\gamma_z}=0.~(b)
\end{equation}
The mathematical expressions for Eqs. (\ref{dg}) are rather lengthy to be
reproduced here. However, factorizing terms that can be canceled out and
defining
\begin{equation}
\label{exy}
\eta_{xy}=\frac{1}{2N}\displaystyle\sum_{\vec{q}}\left[2-\left(\cos{q_{x}a}
+ \cos{q_{y}a}\right)\right]\langle\vert \phi_{q} \vert^{2}\rangle_{0},
\end{equation}
\begin{equation}
\label{ez}
\eta_{z}=\frac{1}{N}\displaystyle\sum_{\vec{q}}(1-\cos{q_{z}c})\langle\vert
\phi_{q}\vert^{2}\rangle_{0},
\end{equation}
we arrive at the following expression
\begin{equation}
\label{minenerg2}\left[ 1- \langle(S_{\vec{r}}^{z})^{2}\rangle_{0}\right]
\left(2J\eta_{xy}e^{-\eta_{xy}}+J_{z}\eta_{z}e^{-\eta_{z}}\right)
=\frac{k_{B}T}{2}.
\end{equation}
It is interesting to notice that  this very same equation is obtained
either from condition (\ref{dg}a) or (\ref{dg}b), meaning that both variational
parameters cannot be obtained from this equation alone. However, from the 
Gaussian variable definitions (\ref{gauss})-(\ref{difangcf}) one can deduce 
the following additional relation for the fluctuations $\eta_{xy}$ and $\eta_z$
\begin{equation}
\label{extraeq}
2\gamma\eta_{xy}+\gamma_z\eta_z=\frac{k_BT}{2}.
\end{equation}
Comparing now Eqs. (\ref{minenerg2}) and (\ref{extraeq}) we end at the
following identifications
\begin{equation}
\label{noacople1b}\gamma=J
\left[ 1- \langle(S_{\vec{r}}^{z})^{2}\rangle_{0}\right]e^{-\eta_{xy}},
\end{equation}
\begin{equation}
\label{noacople2b}\gamma_z=J_z
\left[ 1- \langle(S_{\vec{r}}^{z})^{2}\rangle_{0}\right]e^{-\eta_{z}},
\end{equation}
which are now two parametric equations from which the two variational 
parameters can be obtained. These equations can be put in a more convenient
form by noting that 
\begin{eqnarray}
\label{soma1} e^{-\eta_{xy}}=
\exp{\left\{-\frac{1}{N}\sum_{  \vec{q}}
\frac{k_{B}T(1-\lambda_{{q}})}{2\left[2{\gamma}(1-\lambda_{{q}})
+{\gamma_{z}}(1-\lambda_{{qz}})\right]}\right\}}.
\end{eqnarray}
Taking the continuum limit of the above equation in cylindrical coordinates 
$\frac{1}{N}\displaystyle\sum_{\vec{q}}\to
\frac{a^{2}c}{2\pi^{4}}\int_{0}^{2\pi}d\theta
\int_{0}^{\frac{\pi}{a}}qdq\int_{-\frac{\pi}{c}}^{\frac{\pi}{c}}dq_{z}$
and the long wave length limit $\vec{q}\approx0$  we find
\begin{eqnarray}
\label{resut1}
e^{-\eta_{xy}}=\hskip3.5cm
\nonumber\\ \exp{\left\{-\frac{k_{B}T}{2{\gamma}}
\left[\frac{\arctan{g^{\frac{1}{2}}}}{3g^{\frac{1}{2}}}+
\frac{1}{6}-\frac{g}{6}\ln{(1+\frac{1}{g})}\right]\right\}},
\end{eqnarray}
where $g=\frac{{\gamma_{z}}}{{\gamma}}$. Analogously we find
\begin{eqnarray}
\label{resut2}
e^{-\eta_{z}}=
\exp\{-\frac{k_{B}T}{2\gamma}\hskip2cm \nonumber\\
\times \left[\frac{1}{g}-\frac{1}{3g^{\frac{3}{2}}}\left(
2\arctan{g^{\frac{1}{2}}}
+g^{\frac{1}{2}}-g^{\frac{3}{2}}\ln(1+\frac{1}{g})\right)\right]\},
\end{eqnarray}
so that the variational parameters are finally obtained from
\begin{eqnarray}
\label{noacople3} \gamma&=& J(1- \langle
(S_{\vec{r}}^{z})^{2}\rangle_{o})\exp\{-\frac{k_BT}{2\gamma}
[\frac{\arctan{g^{\frac{1}{2}}}}
      {3g^{\frac{1}{2}}}         \nonumber \\     
&+&\frac{1}{6}
-\frac{g}{6}\ln{(1+\frac{1}{g})]\}},
\end{eqnarray}
\begin{eqnarray}
\label{noacople4} \gamma_z&=& J_{z}(1- \langle
(S_{\vec{r}}^{z})^{2}\rangle_{o})\exp\{-\frac{k_{B}T}{2{\gamma}}
[\frac{1}{g}- \nonumber \\
&-&\frac{1}{3g^{\frac{3}{2}}}(2\arctan{g^{\frac{1}{2}}}
+ g^{\frac{1}{2}}-g^{\frac{3}{2}}\ln{(1+\frac{1}{g})})]\}.
\end{eqnarray}
The two equations above can also be obtained by employing the usual SCHA.
Thus, for a given value of $t=k_BT/J$, $D/J$ and $J_z/J$ one can solve the
non-linear system (\ref{noacople3}) and (\ref{noacople4}) to get 
$\gamma/J$ and $\gamma_z/J$, and from them the desired thermodynamics of 
the model.
For example, by taking the continuum limit in the long wave length regime 
the $x$ component of the magnetization is given by
\begin{eqnarray}
\label{magh} m&=&(1-\frac{1}{2}\langle(S_{\vec{r}}^{z})^{2}
\rangle_{0})\exp\{-\frac{k_{B}T}{2\pi^{2}{\gamma}}
[\frac{\arctan{(g^{\frac{1}{2}})}}
{g^{\frac{1}{2}}} \nonumber \\
&+&\frac{1}{2}\ln{(1+\frac{1}{g}})]\}.
\end{eqnarray}
The transition temperature is obtained when the only solution of Eqs.
(\ref{noacople3}) and (\ref{noacople4}) is $\gamma=\gamma_z=0$.

Before exploring possible simpler solutions of Eqs. (\ref{noacople3})
and (\ref{noacople4}) and presenting the numerical results, it is worthwhile 
now to discuss and compare some limiting cases. 

Let us consider first $D=0$.
In this case we have the anisotropic $XY$ model. The 
out-of-plane spin fluctuation takes the form
\begin{equation}
\label{sm2D0}
\langle(S_{\vec{r}}^{z})^{2}\rangle_{0}=\frac{k_{B}T}{4J+2J_z},
\end{equation}
which is the same as that obtained by Costa et al. \cite{tonico2}.
The transition temperature for 
the two-dimensional model $J_z=0$, as well as the isotropic three-dimensional
model $J_z=J$, are also identical to those from reference \cite{tonico2},
respectively, $t_c=1.076$ and $t_c=1.605$. However, it is worth to stress here
that the present approach not only is a generalization over the model treated 
by Costa et al. \cite{tonico2} to other values of
the axial interaction $J_z$, but also includes, as it will be seen below, 
the important contribution of the crystal field interaction $D$.

In the limit $D\to\infty$, as it will be discussed below, 
we have the planar rotator model. Again, we find the same
transition temperature for the two-dimensional model $t_c=1.472$ and,
for the three-dimensional model, $t_c=2.190$ \cite{tonico2}. 
The transition temperatures for $D=0$ and $D\to\infty$ are given in Table 
\ref{tab1} for the isotropic two- and three-dimensional models together with 
those coming from other approaches. 

\begin{table}
\caption{\it Reduced transition temperatures $t_c$ for the anisotropic $XY$
model in some isotropic limiting cases according to Monte Carlo (MC),
series, present approach, and Assumption 1 and Assumption 2 
(Ass. 1-2).}
\label{tab1}
\vskip0.1in
\begin{tabular}{|c|c|c|c|} \hline
  & MC or series & present & Ass. 1-2 \\ \hline
 \hline
$D\to\infty$, $J_{z}=0$  &  $0.90$\cite{chakra}
           &  1.472     &  1.472  \\ \hline
$D\to\infty$, $J_{z}=J$  &  $2.17$\cite{kohr}
           &  2.190     &  2.207  \\ \hline
$D=0$, $J_{z}=0$          &  $0.78(2)$\cite{kawa}
      &  1.076     &  1.076\\ \hline
$D=0$, $J_{z}=J$  &  $1.54(1)$\cite{tonico2},  $1.55$\cite{ferer}   &  1.605

&  1.613 \\ \hline
\end{tabular}
\end{table}

In addition, the present approximation can also be used to evaluate the
behavior of the $x$ component of the magnetization for small axial
interaction   $J_{z}\ll J$ and any value of $D$. In this limit, we also have
$g\rightarrow 0$ so  Eq. (\ref{magh}) gives
\begin{equation}
\label{mag6} m\approx \left[1-\frac{1}{2}\langle(S_{\vec{r}}^{z})^{2}
\rangle_{0}\right]\left(\frac{{\gamma}_{z}}{{\gamma}}\right)^{\frac{k_{B}T}
{4\pi^{2}{K}}}.
\end{equation}
For the planar rotator model we have $\langle(S_{\vec{r}}^{z})^{2}
\rangle_{0}=0$ and the above equation should be compared to
\begin{equation}
\label{mag5} m\approx
\left(\frac{{\gamma}_{z}}{{\gamma}}\right)^{\frac{k_{B}T}{8\pi {K}}},
\end{equation}
obtained from a SCHA where $\gamma_z/\gamma$ plays the role of
$K_z/K$ of Ref. \cite{tonico1} and
\begin{equation}
\label{mag4} m=\left(\frac{\gamma_{z}}{\gamma}\right)^{\frac{k_{B}T}{8\pi J}},
\end{equation}
obtained from spin wave theory where now $\gamma_z/\gamma$ plays the role of
$J_z/J$ of Ref. \cite{hika}. The exponent here is not the same,
although numerically comparable to other approaches.

\subsection{3. Numerical Results, Parametric Procedure and Simple Assumptions}

\subsection{ Numerical Results}

\begin{figure}[ht]
\includegraphics[clip,angle=0,width=8.5cm]{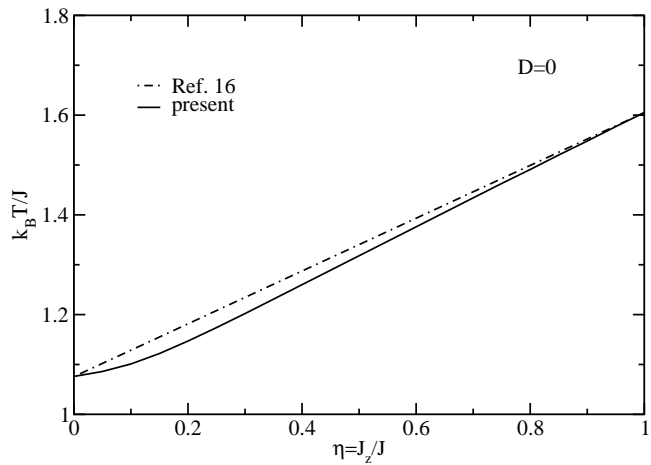} 
\caption{\label{dfD0}Reduced transition temperature $t=k_BT/J$ as a 
function of $\eta=J_z/J$ for $D=0$.
The solid line represents the present approach and 
the dot-dashed line the results from Ref. \cite{tonico2}.
}
\end{figure}

In Figure \ref{dfD0}
we have the reduced transition temperature as a function of the anisotropy
$\eta=J_z/J$ for $D=0$. From hereon the present approach means the results
obtained by solving the non-linear system of equations (\ref{noacople3}) and 
(\ref{noacople4}) numerically (by using standard iterative procedures)
for the variational parameters. 
As can be seen from this figure the results are quite
similar to those from reference \cite{tonico2} in the region close to the
isotropic three-dimensional case $J_z\sim J$.
On the other hand, the slope of the phase boundary close to isotropic 
two-dimensional case $J_z=0$ is zero according to the
present approach while the procedure from reference \cite{tonico2} for the
$XY$ model  (and also for the planar rotator model \cite{tonico1}) furnishes a 
positive slope.
%
%

\begin{figure}[ht]
\includegraphics[clip,angle=0,width=8.5cm]{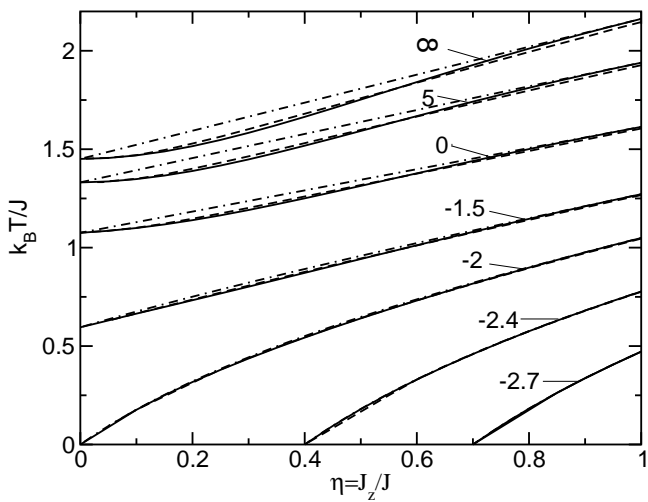} 
\caption{\label{dfD}
Reduced transition temperature $t=k_BT/J$ as a function of 
$\eta=J_z/J$ for several values of $D/J$ (indicated by the numbers)
according to the present approach (full lines), and Assumption 1 
(dashed lines) and Assumption 2 (dot-dashed lines).
}
\end{figure}

Figure \ref{dfD} shows the phase diagram in the reduced temperature $t=k_BT/J$
and $\eta=J_z/J$ plane for several values of $D/J$. 
One can clearly note that the crystalline anisotropy $D$ plays an
important role in the critical behavior of the model. 

For positive values of the crystal field $D>0$ the
transition temperature increases as $D$ increases, since in this case
the out-of-plane fluctuations are reduced implying a greater tendency
of the spin components to lie in the $xy$ plane. For $D\to\infty$ we 
recover the planar rotator model because there will be no $z$ component
of any spin. As shown in Table \ref{tab1}, 
in this limit, the result $t_c=1.472$ for $J_z=0$ are quite different
from that obtained by Monte Carlo simulations $t_c=0.90$, reflecting the 
fact that the present variational approach does not take into account 
vortices effects,
which are relevant for such two-dimensional model. On the other hand,
for $J_z=J$ the result $t_c=2.190$ for the isotropic three-dimensional planar
rotator model are quite comparable to the Monte Carlo simulations $t_c=2.17$.
Moreover, in the 
two-dimensional limit ($J_z=0)$,  Eqs. (\ref{noacople3}) 
and (\ref{noacople4})  can be written as
\begin{equation}
\label{tcJz0}
\frac{k_BT_c}{J}=t_c=\frac{8+4\frac{D}{J}}{2+e(2+\frac{D}{J})},
\end{equation}
which gives the transition temperature of Fig. \ref{dfD} for $\eta=0$ 
down to the value $\frac{D}{J}=-2$. 
%
%

For negative values of the crystal field $D<0$ we have an inverse situation. 
We note
that as $D$ decreases the transition temperature also decreases. This can be
understood because the out-of-plane fluctuations are now enhanced 
implying in a tendency for the spins to lie out of the $xy$ plane and to
become more Ising like. For a given value of $D$ there is a critical
value of the ratio $\eta_c=J_z/J$ at which the temperature goes to zero.
This value is given by 
\begin{equation}
\label{etac}
\eta_c=-D/J-2
\end{equation}
at which the out-of-plane
fluctuation (\ref{sm2}) diverges.

\begin{figure}[ht]
\includegraphics[clip,angle=0,width=8.5cm]{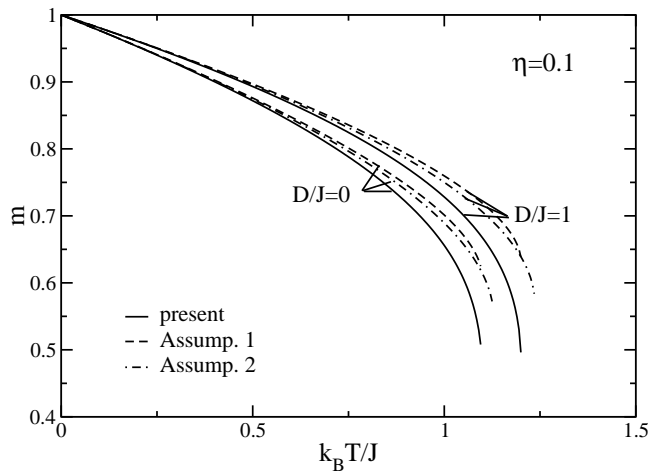} 
\caption{\label{mD01}
Magnetization $m$ as a function of the reduced temperature $k_BT/J$
for $\eta=J_z/J=0.1$ and
two different values of the crystal field, $D/J=0$ and $D/J=1$, according to
the present approach (full lines), and Assumption 1 (dashed lines) and 
Assumption 2 (dot-dashed lines).
}
\end{figure}

\begin{figure}[ht]
\includegraphics[clip,angle=0,width=8.5cm]{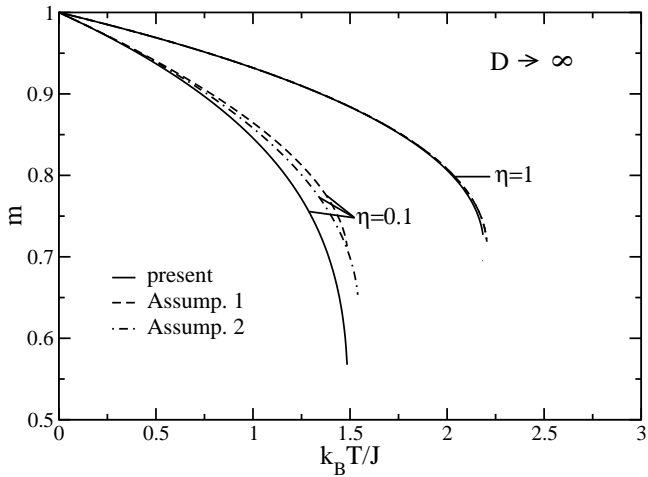} 
\caption{\label{mDinf}
Magnetization $m$ as a function of the reduced temperature $k_BT/J$
for $D\to\infty$ and
two different values of $\eta=J_z/J$, $\eta=0.1$ and $\eta=1$, according to
the present approach (full lines), and Assumption 1 (dashed lines) and 
Assumption 2 (dot-dashed lines).
}
\end{figure}

Figures \ref{mD01} and \ref{mDinf} show the temperature
dependence of the magnetization obtained from Eq. (\ref{magh}) where one can
see a discontinuous behavior.
This discontinuity  in the magnetization
is an artifact of the present method. The transition temperature is
given when the non-linear system of equations
(\ref{noacople3}) and (\ref{noacople4}) admit only the trivial solution.
In all cases they do not go smoothly to the trivial solution presenting
thus a discontinuity. This is a general feature of such methods.
For instance, the temperature behavior for $D\to\infty$ and $J_z/J=0.1$ and
$J_z/J=1$ shown in Figure \ref{mDinf} is quite similar to 
those given in Figure 3 of reference \cite{tonico1}.
%
%

\subsection{Parametric Procedure and Simple Assumptions}

If we take the ratio between the variational stiffness 
$g=\frac{\gamma_z}{\gamma}$,
we can compute the quadratic fluctuations given by Eqs. (\ref{exy}) and 
(\ref{ez}) by taking the continuum limit in the long wavelength regime
\begin{equation}
\label{exyzF}
\eta_{xy}=\frac{k_BT}{2\gamma}F(g),~~~~~~\eta_{z}=\frac{k_BT}{2\gamma}
\left[\frac{1-2F(g)}{g}\right],
\end{equation}
where
\begin{equation}
\label{Fg}
F(g)=\frac{1}{3}\left[\frac{\arctan{(g^{\frac{1}{2}})}}
{g^{\frac{1}{2}}}+ \frac{1}{2}
-\frac{1}{2}g\ln{(1+\frac{1}{g}})\right].
\end{equation}
On the other hand, Eq. (\ref{minenerg2}) can be written as
\begin{equation}
\label{QG}G(Q)=Q\ln{Q}\left[2+\eta\alpha Q^{\alpha-1}\right]=
\frac{-t}{2\left[1-
\langle (S_{\vec{r}}^{z})^{2}\rangle_{0}\right]},
\end{equation}
where $Q=e^{-\eta_{xy}}$, $\alpha=\frac{\eta_z}{\eta_{xy}}$ and 
$\eta=\frac{J_z}{J}$.
From Eqs. (\ref{exyzF}) the parameter $\alpha$ and, in the same way from
Eqs. (\ref{noacople1b}) and (\ref{noacople2b}) the quantity $Q$, can be written
in terms of the ratio $g$ as
\begin{equation}
\label{alpha}\alpha=\left[\frac{1-2F(g)}{2F(g)}\right],~~~~
\label{Q}Q=\left(\frac{g}{\eta}\right)^{\frac{1}{\alpha-1}}.
\end{equation}
Finally, the temperature can be expressed in terms of $G(Q)$ as
\begin{equation}
\label{t}t=\frac{-2G(Q)}{1-\frac{J}{\Omega}G(Q)}.
\end{equation}
So, knowing a priori the Hamiltonian parameters $\eta$ and $D/J$ and for a 
given value
of $g$ one gets 
\begin{equation}
\label{flux}
g \rightarrow F(g) \rightarrow \alpha  \rightarrow Q \rightarrow G(Q)
\rightarrow t. 
\end{equation}
This is indeed what really happens when we use the numerical solution
of the implicit self-consistent equations  (\ref{noacople3}) and 
(\ref{noacople4}).
However, it would be quite nice if one could track the inverse path of Eq.
(\ref{flux}). Unfortunately, as one can see, this is in fact not possible 
since for a given
$t$ (the usual parametric procedure to get the different thermal dependences)
we cannot solve Eq. (\ref{QG}) because $\alpha$ is unknown. It is at this point
where some additional assumptions could be made in order to explore simpler 
solutions of Eq. (\ref{QG}).

\vskip0.5cm
{\bf Assumption 1.} ~~~ $\alpha=\eta$
\vskip0.5cm
In this case Eq. (\ref{QG}) reduces to 
\begin{equation}
\label{Qa1}Q\ln{Q}\left[2+\eta^2 Q^{\eta-1}\right]=
\frac{-t}{2\left[1-
\langle (S_{\vec{r}}^{z})^{2}\rangle_{0}\right]}.
\end{equation}
In Figure \ref{dfD} it is shown, by the dashed lines, the corresponding 
critical
temperatures according to this assumption. It can be seen that, within that 
scale, Assumption 1 and the present variational approach are almost 
indistinguishable. For $D=0$
the transition temperature obtained from Eq. (\ref{Qa1}) for
the two-dimensional model $J_z=0$ is identical to that from the present one
$t_c=1.076$. However, for the isotropic three-dimensional model $J_z=J$
one has $t_c=1.613$, which is slightly different from
the value $t_c=1.605$ according to reference \cite{tonico2}.
On the other hand, in the limit $D\to\infty$ we find the same transition 
temperature for the two-dimensional planar rotator model $t_c=1.472$ and,
for the three-dimensional case,
$t_c=2.207$  is comparable to
$t_c=2.190$ obtained from Ref. \cite{tonico2}. These values are
depicted in Table \ref{tab1} together with those coming from other
approaches. The  magnetization as a function of the temperature
is presented in Figures
\ref{mD01} and \ref{mDinf}. A different behavior is observed, except for the 
planar rotator model.

The quite good agreement of the present approach and Assumption 1 reflects 
the fact that
the term $\eta\alpha Q^{\alpha-1}$ in Eq. (\ref{QG})
varies very weakly with $g$ and thus with
temperature. Reproducing the correct limits at $\eta=0$ and $1$ the phase 
boundaries are then a sort of smooth interpolation for different values of the
anisotropy, explaining the accidental good agreement. With this in mind other
simplifications can also be done.
\vskip0.5cm
{\bf Assumption 2.}~~~$F(g)=\frac{1}{2+g}$
\vskip0.5cm
In the interval $[0,1]$, the function $F(g)$ given by Eq. (\ref{Fg}) can be 
approximated by (with an error of order less than $10\%$)
\begin{equation}
\label{modF}F(g)=\frac{1}{2+g}.
\end{equation}
This implies that $\alpha=1$ and $\eta_z=\eta_{xy}$, meaning that the quadratic
fluctuations are not very different in vertical or in-plane bonds. The 
self-consistent Eq. (\ref{QG}) adopts now a very simple form
\begin{equation}
\label{Qa2}Q\ln{Q}\left[2+\eta\right]=
\frac{-t}{2\left[1-
\langle (S_{\vec{r}}^{z})^{2}\rangle_{0}\right]}.
\end{equation}
Unlike Eq. (\ref{Qa1}) the above equation permits obtaining an analytical 
expression for the critical
temperature. Since $\frac{dQ}{dt}=0$ at $t_c$ one gets $Q_c=e^{-1}$ and
\begin{equation}
\label{tca2}t_c=\frac{2(2+\eta)}{\frac{J}{\Omega}(2+\eta)+e}.
\end{equation}
The above equation also yields  exactly the same limiting results as in the
last column of Table \ref{tab1}. The boundaries as a function of $D$ are given
in Figure \ref{dfD} by the dot-dashed lines. In this case, for positive values
of the crystal anisotropy the agreement is not as good as for negative values. 
The  corresponding magnetizations as a function of the temperature
are also shown in Figures \ref{mD01} and \ref{mDinf}.

\section{ 4. concluding remarks}

The anisotropic $XY$ model in a crystalline field has been studied according
to a variational approach for the free energy.
This system is a generalization of the planar rotator model and the 
anisotropic $XY$ model previously treated in the literature
\cite{tonico1,tonico2}. We believe we have obtained a satisfactory 
picture of the thermodynamic behavior of the model as a function of its 
parameters. Not only is the phase diagram in the anisotropic case 
$J_z\ne J$ different from the previous one
obtained for $D=0$, but the crystal field has been show to play an important 
role in the critical behavior of the system.

Regarding Assumptions 1 and 2, one could see that the former one
($\alpha=\eta$) would imply that the mean quadratic fluctuation on a given 
bond scales with the coupling constant of the bond,  while the latter one 
($\alpha=1,~\eta=1$) means
that the quadratic fluctuations are not very different in vertical or in-plane
bonds. So, according to Assumption 2, the average contribution to
the internal energy from each bond scales approximately with the coupling
constant of the considered bond. This seems more acceptable than the 
conceptually wrong assumption $\alpha=\eta$, since one expects that softer
bonds should develop stronger fluctuations. Thus, at first sight, from the
numerical point of view, the better agreement of Assumption 1 with 
the complete approach can be ascribed to an artifact of the method.

\begin{figure}[ht]
\includegraphics[clip,angle=-90,width=9.3cm]{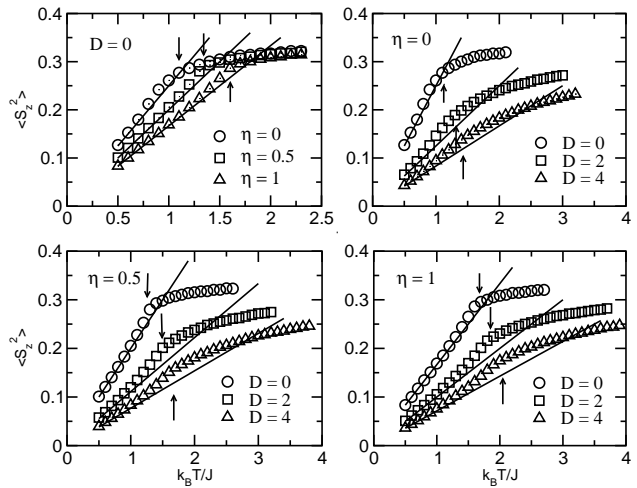} 
\caption{\label{val}
Out-of-plane spin fluctuation as a function of the temperature for 
several values
of the Hamiltonian parameters. The full lines are the results from Eq.
(\ref{sm2}) and the different symbols are Monte Carlo simulations \cite{rod}.
The arrows indicate the transition temperatures obtained from the 
present approximation. The MC results saturates at $1/3$ as 
$T\to\infty$.
}
\end{figure}

It is clear from the approximation employed in this work that it
should be valid only at low temperatures. It is also rather surprising 
that the values for the transition temperatures shown in Table \ref{tab1} for
the three-dimensional model are quite comparable to those coming from more
reliable methods (for the two dimensional model one would not expect such
agreement due to the vortices effects). It would then be quite nice to have
a clearer picture of the range of the temperature validity of the present 
procedure. In Figure \ref{val} we show the out-of-plane spin fluctuations given
by Eq.(\ref{sm2}) for several values of the Hamiltonian parameters compared
to Monte Carlo simulations. We have chosen this quantity because it is easily
obtained from the present approximation, Eq.(\ref{sm2}). 
The simulations have been done in 
the three-dimensional model described by Hamiltonian (\ref{ham}) 
and employing the single spin-flip Metropolis algorithm. A finite 
lattice of $L\times L\times L$ ($L=14$)
has been used with periodic boundary conditions. The averages have been taken
by considering $10^5$ Monte Carlo steps (MCS) per spin after equilibration. 
For the  first temperature, $100L^2$ MCS have been discarded and,
with the previous configuration being the initial configuration for the next 
temperature, additional $3\times 10^3$ have been discarded. 
It is evident from this figure that the
transition temperatures are not far from the temperature range where
the corresponding results are comparable to the Monte Carlo simulations.
%
%

{\bf Acknowledgments - }The authors would like to thank CNPq, FAPEMIG and 
CAPES (Brazilian agencies) for financial support. Special thanks for the
nicely and extensively written referee comments and suggestions are also 
addressed.


\begin{thebibliography}{00}

\bibitem{kobe} S. Kobe, Braz. J. Phys. {\bf 30}, 649 (2000).
\bibitem{stinch} R. B. Stinchcombe, in {\it Phase transitions and Critical
 Phenomena}, edited by C. Domb and J. L. Lebowitz, vol. 7 (London, Academic, 
 1983).
\bibitem{belan} D. P. Belanger, Braz. J. Phys. {\bf 30}, 682 (2000).
\bibitem{wolf} W. P. Wolf, Braz. J. Phys. {\bf 30}, 794 (2000).
\bibitem{be}V. L. Berezinskii, Sov. Phys. JETP {\bf 32}, 493 (1970). 
\bibitem{kt}J. M. Kosterlitz and D. J. Thouless, J. Phys. C {\bf 6},  
1181 (1973).
\bibitem{svistov} L. E. Svistov et al. cond-mat/0603617.
\bibitem{liu} S. A. Liu, Q. Wang, H. Liu, T. L. Chen, Commun. Theor. Phys. 
 {\bf 32}, 339 (1999).
\bibitem{pierson}S. W. Pierson, Phys. Rev. B {\bf 51}, 6663 (1994).
\bibitem{choi}M. S. Choi and S. Lee, Phys. Rev. B {\bf 51}, 6680 (1995).
\bibitem{franz} M. Franz and A. P. Iyengar, Phys. Rev. Lett. {\bf 96},
 047007 (2006).
\bibitem{lms} E. Lieb, T. Schultz and D. Mattis, Ann. Phys. (NY) {\bf 16},
 407 (1961).
\bibitem{beth}S. L. Menezes, M. E. Gouvea and A. S. T. Pires, Phys. Lett. A
 {\bf 166}, 330 (1992).
\bibitem{tonico3}A. S. T. Pires, Sol. State Comm. {\bf 100}, 791 (1996). 
\bibitem{tonico1}A. R. Pereira, A. S. T. Pires and M. E. Gouvea, Phys. Rev. B
 {\bf 22}, 16413 (1995).
\bibitem{tonico2} B. V. Costa, A. R. Pereira and A. S. T. Pires
 Phys. Rev. B {\bf 54}, 3019 (1996).
\bibitem{landau} D. P. Landau and K. Binder, {\it A Guide to Monte Carlo 
 Simulations in Statistical Physics}, Cambridge (2000).
\bibitem{sach} Subir Sachdev in {\it Quantum Phase Transitions}, Cambridge
 University Press (Cambridge, 1999).
\bibitem{lonz}G. L. Lonzarich, Nature Phys. {\bf 1}, 11 (2005). 
\bibitem{min}P. Minnhagen and P. Olsson, Phys. Rev. B{\bf 44}, 4503 (1991).
\bibitem{shenoy} S. R. Shenoy and B. Chattopadhyay, Phys. Rev. B {\bf 51},
 9129 (1995).
\bibitem{eric} E. Roddick and D. Stroud, Phys. Rev. Lett. {\bf 74}, 1430
 (1995).
\bibitem{lozo}Y. E. Lozovik and S. G. Akopov, J. Phys. C {\bf 14}, L31 
 (1981).
\bibitem{wood}D. M. Wood and D. Strout,Phys. Rev. B {\bf 25}, 1600 (1982).
\bibitem{fish1}R. S. Fishman and D. Strout, Phys. Rev. B {\bf 38}, 290 (1988).
\bibitem{fish2}  R. S. Fishman, Phys. Rev. B {\bf 38}, 11996 (1988).
\bibitem{ari2}D. Ariosa et al., J. Phys. I(France) {\bf 51}, 1373 (1990).
\bibitem{ari1}D. Ariosa and H. Beck, Phys. Rev. B {\bf 43}, 344 (1991).  
\bibitem{ari3}D. Ariosa and H. Beck, Helv. Phys. Acta {\bf 65}, 499 (1992).
\bibitem{spis}D. Spisak Physica B {\bf 190}, 407 (1993).
\bibitem{darcy}L. M. Castro, A. S. T. Pires and J. A. Plascak, J. Mag.
 Mag. Mat. {\bf 248}, 62 (2002).
\bibitem{chakra}S. Chakravarty, G. L. Ingold, S. Kivelson and A. Luther,
 Phys. Rev. Lett. {\bf 56}, 2303 (1986).
\bibitem{kohr}G. Kohring, R. E. Shrock and P. Wills, Phys. Rev. Lett. {\bf 57},
 1358 (1986).
\bibitem{kawa}C. Kawabata and A. R. Bishop, Sol. Stat. Comm. {\bf 60}, 169
 (1986).
\bibitem{ferer} M. Ferer, M. Moore and M. Wortis, Phys. Rev. B   {\bf 8}, 5205
 (1973).
\bibitem{hika}S. Hikami and T. Tsuneto, Prog. Theor. Phys. {\bf 63}, 387 (1980)
\bibitem{rod}R. S. T. Freire, private communication. 

\end{thebibliography}
\end{document}